# Title

Real-time convolutional voltammetry enhanced by energetic (hot) electrons and holes on a surface plasmon waveguide electrode

## Authors

Zohreh Hirbodvash,[1] Elena A. Baranova,[2,3] and Pierre Berini[1,4,*]//

## Affiliations

[1]Dept. of Physics, University of Ottawa, 150 Louis Pasteur, Ottawa, Ontario, K1N 6N5, Canada
[2]Department of Chemical and Biological Engineering, University of Ottawa, 161 Louis-Pasteur, Ottawa, ON K1N 6N5, Canada
[3]Centre for Catalysis Research and Innovation, University of Ottawa, 161 Louis-Pasteur, Ottawa, ON K1N 6N5, Canada
[4]School of Electrical Engineering and Computer Science, University of Ottawa, 800 King Edward Ave., Ottawa, Ontario, K1N 6N5, Canada
*Corresponding Author; berini@eecs.uottawa.ca

## Keywords

## Abstract

Surface plasmon polaritons (SPPs) propagating along a waveguide working electrode are sensitive to changes in local refractive index, which follow changes in the concentration of reduced and oxidised species near the working electrode. The real-time response of the output optical power from a waveguide working electrode is proportional to the time convolution of the electrochemical current density, precluding the need to compute the latter *a posteriori* via numerical integration. The theoretical optical response of a waveguide working electrode is derived, and validated experimentally via chronoamperometry and cyclic voltammetry measurements under low power SPP excitation, for various concentrations of potassium ferricyanide in potassium nitrate electrolyte at various scan rates. Increasing the SPP power induces a regime where the SPPs no longer act solely as a probe of electrochemical activity, but also as a pump creating energetic electrons and holes via absorption in the working electrode. In this regime the transfer of energetic carriers (electrons and holes) to the redox species dominates the electrochemical current density, which becomes significantly enhanced relative to equilibrium (low SPP power) conditions. In this regime the output optical power remains proportional to the time convolution of the current density, even with the latter significantly enhanced by the transfer of energetic carriers.

## Introduction

Convolutional voltammetry, developed initially by Imbeaux and Saveant [1], is based on an electrochemical technique, such as voltammetry, chronoamperometry, or chronocoulometry, followed by a mathematical transformation (convolution) of measured signals [2-5]. Convolutional voltammetry is useful because the results are directly related to the concentration of electroactive species at the working electrode surface. Conversely, voltammetric, chronoamperometric, or chronocoulometric measurements are directly related to the flux of a species. Thus, convolutional voltammetry yields complementary results and, furthermore, offers advantages over conventional electrochemical techniques, such as low sensitivity to a drop in current passing through the electrochemical system [2,6], and independence from the scan rate in cyclic voltammetry [7]. Convolutional voltammetry has been used to determine the diffusion constant, bulk concentration, and the number of transferred electrons for an electroactive species of interest. However, convolved signals are not generated directly in a conventional electrochemical experiment, so convolutional techniques are deemed less convenient [8].

A metal-coated prism enabling surface plasmon resonance (SPR) measurements in the Kretschmann configuration can be placed in an electrochemical cell, and the metal surface used to support surface plasmon polaritons (SPPs) can operate simultaneously as the working electrode (WE), enabling joint electrochemical SPR measurements (EC-SPR) [9-11]. The SPR response in such systems directly monitors changes in the optical parameters of the redox species at the metal/electrolyte interface as electrochemical reactions take place. In such configurations, the SPR signal measured in real-time during an electrochemical reaction follows the convolution of the electrochemical current [11]. EC-SPR can thus be used as a platform for measuring convolutional voltammetry directly and quantitatively [9-11].

EC-SPR voltammetry techniques are similar to techniques involving a microdisk electrode or a rotating disk electrode, in that they are scan rate independent [12]. In conventional electrolytes (*e.g.*, salts in molecular solvents), steady-state microdisk electrode or rotating disk electrode voltammetry is effective in determining fundamental parameters, *e.g.*, diffusion coefficients or bulk concentrations of electroactive species [12], but have proven difficult to apply in highly viscous media such as ionic liquids at room temperature. Several studies have demonstrated that convolutional voltammetry is robust enough to determine accurate values of diffusivity, bulk concentration, and the stoichiometric number of electrons for electrode reactions in molecular solvents and room temperature ionic liquids [12].

Electrochemical electrodes formed into plasmonic structures are also being used in studies of plasmon-induced energetic (hot) carrier electrochemistry [13]. Advantageously, by separating oxidation and reduction reactions electrochemically, the role of energetic holes can be distinguished from that of energetic electrons. Energetic carriers are created in a WE via SPP absorption therein [14], but absorption also raises the temperature of the metal and heat diffuses into the surrounding reaction volume [15-17].

It is not trivial to separate the effects of temperature from those of energetic carriers, given that electrochemical reactions and fluid dynamics are temperature dependent, but it is possible to isolate effects through careful experimental design and control experiments [18-23].

In recent work, infrared SPPs propagating along a Au WE shaped as a stripe waveguide, induced significant increases in redox currents (10×) along with significant decreases in redox potentials, attributed to the creation of energetic carriers (electrons and holes) by SPP absorption in the Au stripe [23]. Here we investigate convolutional voltammetry in this system by monitoring the electrochemical current and the output optical power in real time *vs.* the applied potential, as chronoamperometry and cyclic voltammetry (CV) measurements are performed. Output optical power voltammograms indicate that changes in optical signal are caused mainly by changes in the refractive index near the WE, following changes in concentration of the reduced and oxidised forms of the redox species. We also find that the output optical power is proportional to the convolution of the electrochemical current in the low SPP power regime, and that this relationship persists at high SPP powers where the electrochemical current is dominated by redox reactions involving energetic carriers. We formulate a theoretical model of the system which we validate experimentally.

**Theoretical**

The system of interest exploits Bloch long-range SPPs [24] propagating along a metal stripe waveguide operating simultaneously as a WE, coupled to external optoelectronics using input and output grating couplers [25], as sketched in Fig. 1. SPP transmission is monitored by measuring the output optical power, $P_{out}$, which is the optical measurand. The current flowing through the WE into the redox species is the electrical measurand. Advantages of this system include a compact arrangement, a WE that is defined lithographically and is well-controlled structurally, SPPs that propagate over the entire length of the WE with complete overlap, and a thin WE which enhances the escape probability of energetic carriers created therein via SPP absorption [23].

The expression that relates the output optical power $P_{out}$ to the incident optical power $P_{inc}$ is written as:

$$P_{out}(n_c) = P_{inc} T_{in}(n_c) e^{-2\alpha(n_c)l_3} T_{out}(n_c) = P_{inc} T_s(n_c) \qquad (1)$$

where $T_{in}$ is the transmittance of the incident beam to SPPs propagating along the WE coupled by the input grating, $T_{out}$ is the transmittance of the SPPs to the output beam coupled by the output grating, $\alpha$ [m$^{-1}$] is the field attenuation coefficient of the propagating SPPs, and $l_3$ [m] is the propagation distance of SPPs between the input and output grating couplers (Fig. 1(a)). $T_{in,out}$ and $\alpha$ depend on the refractive index of the electrochemical solution, $n_c$. $T_s$ in the second equality lumps these contributions together as the channel transmittance.

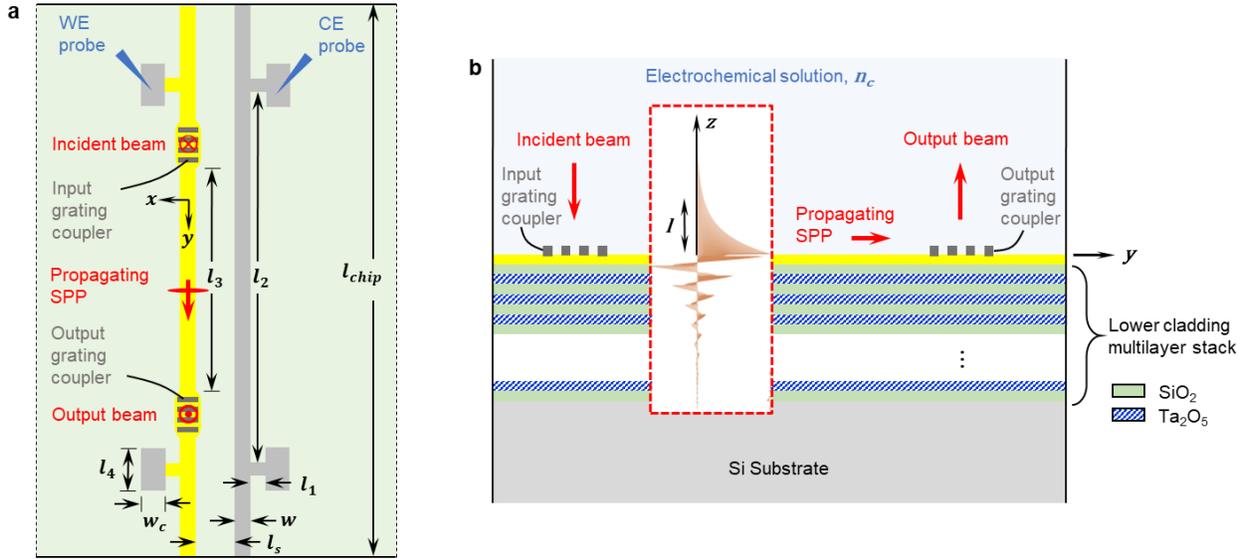

Fig. 1. (a) Schematic in top view of the sensing chip. The thickness of the Au WE (yellow) and Pt CE (gray) is $t = 35$ nm and their dimensions are: $l_1 = 29$ μm, $l_2 = 2600$ μm, $l_3 = 1850$ μm, $l_4 = 250$ μm, $l_{chip} = 3000$ μm, $w_c = 100$ μm, $w = 5$ μm, $l_s = 40$ μm. Experimental scheme: Laser light delivered by a PM-SMF is normally-incident as the input beam on the input grating coupler, exciting SPPs that propagate along the WE. The output grating coupler converts the SPPs to the output beam emerging normally from the chip, captured by a MMF, and measured using a power meter. The WE and CE are contacted using external probes to form a 3-electrode electrochemical system with a Ag/AgCl reference electrode (not shown). The chip is fixed to the bottom of a petri dish immersed by the redox species in electrolyte. (b) Sketch of a longitudinal cross-section taken along the centre of a WE. The electrochemical solution of refractive index $n_c$ covers the structure. The inset shows the computed distribution of Re{$E_z$} of the Bloch long-range SPP that propagates along the WE, showing the 1/e field penetration depth into the electrochemical solution, $l$.

The structure operates optically as an attenuation-based sensor [26, 27], transducing changes in $n_c$ to changes in output optical power, where $n_c$ is the refractive index of the electrochemical solution that covers the stripe. Small changes in output optical power (power response), $\Delta P_{out}$, caused by perturbations in the refractive index, $\Delta n_c$, can be expressed as:

$$\Delta P_{out} = \frac{\partial P_{out}}{\partial n_c} \Delta n_c = S_{B,SPP} \Delta n_c \qquad (2)$$

In the above, $S_{B,SPP}$ [W/RIU] is the bulk sensitivity, which can be determined via calibration or modelling (*cf.* [28]).

Following [11], we consider a reversible redox reaction occurring along one dimension normal to the WE ($z$ axis) and assume that the redox species does not bind or accumulate on the WE but only changes the refractive index of the electrochemical solution in its vicinity. We ignore the double-layer capacitance and the associated charging and discharging currents, as they are usually small in comparison.

The electrochemical solution is comprised of the electrolyte and the redox species. Its time-dependant refractive index in the direction normal to the WE, $z$, is written:

$$n_c(z,t) = n_e + \alpha_O C_O(z,t) + \alpha_R C_R(z,t) = n_e + \Delta n_c(z,t) \quad (3)$$

where $n_e$ is the refractive index of the electrolyte only (constant), $C_O(z,t)$ and $C_R(z,t)$ [mol/m$^3$ or M] are the time-dependant concentration distributions of the oxidised and reduced species, and $\alpha_O = \partial n_e/\partial C_O$ and $\alpha_R = \partial n_e/\partial C_R$ [m$^3$/mol or M$^{-1}$] are the changes in refractive index per unit concentration of oxidised and reduced species (index increments).

The refractive index perturbation is position and time dependant, $\Delta n_c(z,t)$, so the time-dependant power response following Eq. (2), is approximated analytically as the overlap of the normalised decay profile of the SPP intensity over $\Delta n_c(z,t)$:

$$\Delta P_{out}(t) = S_{B,SPP} \left(\frac{2}{l}\right) \int_0^\infty \Delta n_c(z,t) e^{-2z/l} \, dz \quad (4)$$

The exponential term in Eq. (4) models the decay of the SPP intensity from the surface of the metal stripe ($z = 0$) into the solution. The SPP intensity is equal to Re$\{S_y\}/2$, where $S_y$ is the component of the Poynting vector in the direction of SPP propagation ($y$). The $1/e$ field decay length of the SPPs on a metal stripe waveguide into the bounding medium, $l$, depends on the confinement provided by the stripe [29]. In the case of our Bloch long-range SPP, the field decay length into the electrochemical solution is $l \sim 2$ µm, as determined from the mode field distribution computed at $\lambda_0 = 1310$ nm (*cf.* [23]) and plotted as the inset to Fig. 1(b). The SPP intensity therefore drops to $1/e$ at $l/2 \sim 1$ µm.

Using Eq. (3), the power response of Eq. (4) becomes:

$$\Delta P_{out}(t) = S_{B,SPP} \left(\frac{2}{l}\right) \int_0^\infty [\alpha_O C_O(z,t) + \alpha_R C_R(z,t)] e^{-2z/l} \, dz \quad (5)$$

$l$ is a weak function of $C_O(z,t)$ and $C_R(z,t)$, so $l$ is treated as a constant.

The diffusion times over the SPP intensity decay length, $(l/2)^2/(2D_{O,R})$, where $D_O$ and $D_R$ [m$^2$/s] are the diffusion coefficients of the oxidised and reduced species, respectively, are less than a millisecond if the diffusion coefficients are in the range of $10^{-9}$ to $10^{-11}$ m$^2$/s, which is the case for many redox species [30]. If the power measurements (samples) are taken at time intervals that are much longer than the diffusion times, then the instantaneous response implied by Eq. (4) holds.

The time-dependant concentration distributions of the oxidised and reduced species can be obtained by solving Fick's second law of diffusion, written for the redox concentrations, $C_{O,R}(z,t)$:

$$\frac{\partial C_O(z,t)}{\partial t} = D_O \frac{\partial^2 C_O(z,t)}{\partial z^2} \quad \text{and} \quad \frac{\partial C_R(z,t)}{\partial t} = D_R \frac{\partial^2 C_R(z,t)}{\partial z^2} \tag{6}$$

The concentration distribution of the oxidised species, $C_O(z,t)$, satisfying the first of Eqs. (6) is written [31]:

$$\frac{C_O(z,t)}{C_O^b} = 1 - \text{erfc}\left(\frac{z}{\sqrt{4D_O t}}\right) + \exp(\mu z + \mu^2 D_O t) \cdot \text{erfc}\left(\frac{z}{\sqrt{4D_O t}} + \mu\sqrt{D_O t}\right) \tag{7}$$

where μ is:

$$\mu = \frac{k^0}{D_O} \exp\left(\frac{|\eta|}{b}\right) \tag{8}$$

$b = RT/F$ where $R$ is the gas constant (J/(mol·K)), $T$ is the temperature (K), and $F$ is the Faraday constant (C/mol). $\eta$ is the overpotential (V), $\eta = E - E^0$, where $E$ is the potential of the WE through which the current flows, and $E^0$ is the equilibrium potential established when no current flows. $k^0$ is the rate constant at zero potential (m/s). $C_O^b = C_O(z, t = 0)$ is the initial concentration distribution of the oxidised species. The concentration distribution of the reduced species $C_R(z,t)$ follows a similar form.

Fig. 2 plots the computed concentration profiles for 0.5 mM potassium ferricyanide in 100 mM potassium nitrate electrolyte, corresponding to our experimental system of interest, as a function of $z$, the distance from the WE. Following [31], $E$ was set to 0.3 V in the forward direction, corresponding to the oxidation peak. The equilibrium potential is $E^0 = 0.24$ V (other data can be found in [31]). The concentration profiles, plotted $t = 6$ s after application of the potential, indicate a high concentration of the oxidised form on the WE ($z = 0$), along with a low concentration of the reduced form, as expected after application of the (peak) oxidising potential.

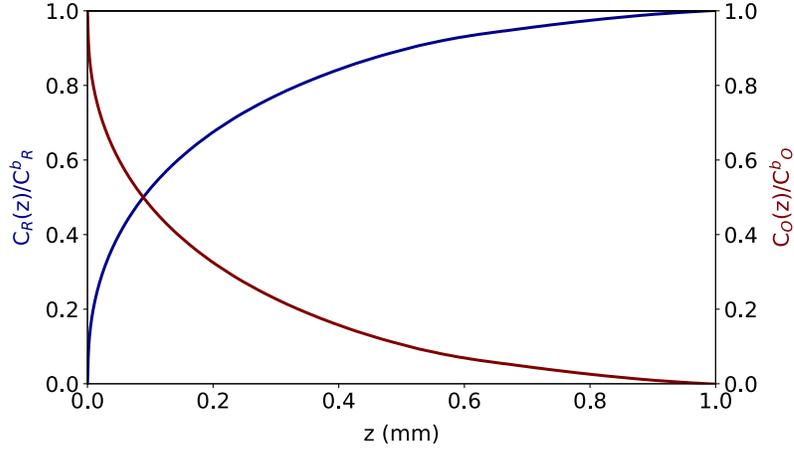

Fig. 2. Computed concentration profiles of potassium ferricyanide as a function of $z$, the distance from the WE, $t = 6$ s after application of the (peak) oxidising potential (0.3V vs. Ag/AgCl reference electrode).

As it is evident from Fig. 2, the concentration profiles reach $\sim C_{O,R}^b$ in a characteristic distance of several hundred microns from electrode surface. Thus, $C_O(z,t)$ and $C_R(z,t)$ do not vary rapidly over $z = l/2 \sim 1$ µm, the characteristic decay length of our SPP intensity, so Eq. (5) can be simplified to:

$$\Delta P_{out}(t) = S_{B,SPP}[\alpha_O C_O(0,t) + \alpha_R C_R(0,t)] \tag{9}$$

where $C_O(0,t)$ and $C_R(0,t)$, are the concentration of the oxidised and reduced species on the WE.

Based on Eq. (9), a relationship between $C_{O,R}(0,t)$ and the electrochemical current is needed to connect the optical power response to the electrochemical current response. Fick's second law of diffusion (Eqs. (6)), can be used for this purpose. The current density flowing through the WE in an electrochemical experiment, $j(t)$, is related to the normal derivative (gradient) of $C_{O,R}(0,t)$ (11):

$$j(t) = nFD_O \left.\frac{\partial C_O(z,t)}{\partial z}\right|_{z=0} = -nFD_R \left.\frac{\partial C_R(z,t)}{\partial z}\right|_{z=0} \tag{10}$$

where $n$ is number of electrons transferred per redox reaction. Using Laplace transforms (cf. [11]), Eqs. (6) are solved subject to the boundary conditions of Eqs. (10), leading to:

$$C_O(0,t) = C_O^0 - \left[nF(\pi D_O)^{1/2}\right]^{-1} \int_0^t j(t')\,(t-t')^{-1/2}dt' \tag{11}$$

$$C_R(0,t) = C_R^0 + \left[nF(\pi D_R)^{1/2}\right]^{-1} \int_0^t j(t')\,(t-t')^{-1/2}dt' \tag{12}$$

where $C_O^0 = C_O(0,0)$ and $C_R^0 = C_R(0,0)$ are the initial concentrations of redox species at the electrode surface.

p. 7

The relationship between the power response and the current density can then be obtained by combining Eq. (9) with Eqs. (11) and (12):

$$\Delta P_{out}(t) = S_{B,SPP}(\alpha_O C_O^0 + \alpha_R C_R^0) + S_{B,SPP}(\alpha_R D_R^{-1/2} - \alpha_O D_O^{-1/2})(nF\pi^{1/2})^{-1} \int_0^t j(t')\,(t-t')^{-1/2} dt'$$

(13)

The optical measurand is $P_{out}(t)$ which can be written with the aid of Eq. (13) as:

$$P_{out}(t) = P_{out}(0) + S_{B,SPP}(\alpha_R D_R^{-1/2} - \alpha_O D_O^{-1/2})(nF\pi^{1/2})^{-1} \int_0^t j(t')\,(t-t')^{-1/2} dt' \quad (14)$$

where $P_{out}(0)$ is the initial output power measured before an electrochemical potential is applied:

$$P_{out}(0) = P_{out}(n_e) + S_{B,SPP}(\alpha_O C_O^0 + \alpha_R C_R^0) \quad (15)$$

$P_{out}(n_e)$ in the above is the output power measured for the electrolyte only. The second term in Eq. (15) originates from Eq. (13) and represents the change in output power due to the addition of the redox species to the electrolyte, but before application of any electrochemical potential. $P_{out}(0)$ thus depends explicitly on the initial concentrations of the redox species, $C_{O,R}^0$. The second term in Eq. (14) is time-dependant and reveals that the output power evolves in time as the convolution of the electrochemical current density $j(t)$, similarly to the resonance angle in an SPR set-up [11].

In attenuation-based sensors [26], the bulk sensitivity, $S_{B,SPP}$, is dependant on the incident optical power, $P_{inc}$, such that increasing the incident power increases the sensitivity. To show this explicitly, we apply its definition (Eq. (2)) to the second form of Eq. (1):

$$S_{B,SPP} = \frac{\partial P_{out}}{\partial n_c} = P_{inc} \frac{\partial T_s}{\partial n_c} = P_{inc} S_{B,T_s} \quad (16)$$

$S_{B,T_s}$ (RIU$^{-1}$) is the power-independent bulk sensitivity of the sensing channel, which can be optimised by design. Substituting Eq. (16) into Eq. (14), we re-write the output power as:

$$P_{out}(t) = P_{out}(0) + P_{inc} S_{B,T_s}(\alpha_R D_R^{-1/2} - \alpha_O D_O^{-1/2})(nF\pi^{1/2})^{-1} \int_0^t j(t')\,(t-t')^{-1/2} dt' \quad (17)$$

From the above, we note that the output power scales linearly with $P_{inc}$, which is controlled independently from the electrochemical variables. This is a distinguishing feature from angular-interrogated EC-SPR systems [11].

Isolating for $S_{B,T_s}$ in the above yields:

$$S_{B,T_s} = \frac{(P_{out}(t)-P_{out}(0))}{P_{inc}} \cdot \frac{1}{(\alpha_R D_R^{-1/2}-\alpha_O D_O^{-1/2})(nF\pi^{1/2})^{-1}\int_0^t j(t')(t-t')^{-1/2}dt'} \tag{18}$$

Thus, $S_{B,T_s}$ can be determined by inserting the optical and electrical measurands, $P_{out}(t)$ and $j(t)$, into Eq. (18). $S_{B,T_s}$ is optimised by design about an electrolyte refractive index, $n_e$, and is independent of the redox species and its concentration, of the electrochemical technique applied, and of any associated electrochemical variables (*e.g.*, scan rate, step potential).

**Experimental**

*A. Set-up*

As sketched in Fig. 1(a), the sensing chip comprises a Au stripe SPP waveguide operating simultaneously as a WE, bearing input and output optical grating couplers, next to a Pt stripe used as the counter electrode (CE). The structures were fabricated on a Si wafer bearing a multilayer dielectric stack consisting of 15-periods of $SiO_2$ and $Ta_2O_5$, as sketched in Fig. 1(b). The dimensions of the structures, given in the caption to Fig. 1(a), and the design of the multilayer stack, were selected such that SPPs propagate along the Au stripe as Bloch long-range SPPs at $\lambda_0 \sim 1310$ nm. These modes propagate with fields confined in the 2D plane normal to the direction of propagation (*y*) and that decay exponentially away from the stripe, penetrating the electrolyte a distance $l \sim 2$ μm, thereby probing the region nearest the stripe. Thick Pt/Cu electrical contact pads were also fabricated to facilitate electrical probing of the stripes. Further details on fabrication and the optical operation of the waveguides can be found in previous reports [24, 25, 32].

A chip was fixed to the bottom of a petri dish and submerged by the redox species in electrolyte. Electrochemical reactions were driven from the WE and CE on chip, using two tungsten needles coated with PMMA (except for their tips), to prevent any electrochemical interference during the measurements. The probing needles were attached to the arms of two positioners and used to probe the contact pads of a Au WE and a Pt CE directly. The Au WE, Pt CE and a Ag/AgCl reference electrode (double junction PH combination, glass body, BNC connector, Sigma-Aldrich Canada Ltd) form a three-electrode system, driven by a potentiostat (WaveDriver20, Pine Research Instrumentation Inc.) to perform cyclic voltammetry (CV) and chronoamperometry measurements.

The experimental scheme made use of a cleaved bow-tie style polarisation-maintaining single-mode optical fibre (PM-SMF) of core diameter 6.6 μm, aligned perpendicularly to the input grating coupler to launch the incident beam that excites SPPs along the WE. The output grating coupler converts the SPPs to the output beam emerging normally from the chip, which was captured by a large-core (200 μm diameter) multi-mode fibre (MMF). Both fibers were mounted to 90° metallic holders on multi-axis micro-positioners. A tunable laser (Agilent 8164A) was coupled to the input of the PM-SMF, and a power meter (PM 100USB, Thorlabs) was used to capture power emerging from the MMF, such that the power transmitted by SPPs along the WE could be continuously monitored *in-situ*. The wavelength of operation ($\lambda_0 = 1350$ nm) was determined by maximising the power transmitted along a submerged WE while tuning the laser wavelength.

*B. Materials - Redox species and electrolyte*

Potassium ferricyanide ($K_3[Fe(CN)_6]$) and potassium nitrate ($KNO_3$) were used as supplied (both >99% purity, Sigma-Aldrich), as the redox species and electrolyte, respectively. The electrolyte covers the WE and CE, and as such also serves as the optical upper cladding through which SPPs on the WE propagate. In order for Bloch long-range SPPs to be supported on the WE [24], glycerol was added to the electrolyte to adjust its refractive index to $n_e = 1.3323$ (at $\lambda_0 = 1312$ nm) as measured using a prism coupler (Metricon) - adding a small amount of glycerol has no effect on the electrochemical response of our redox couple, as verified in previous work [23]. Specifically, 0.2928 g of glycerol (99.5% purity, Sigma-Aldrich) was added per 20 ml of electrolyte, the latter consisting of $KNO_3$ in distilled deionised $H_2O$ at a concentration of 100 mM. Thus, the electrolyte solution used throughout our study consisted of 100 mM $KNO_3$ + 158.5 mM glycerol, into which the redox species $K_3[Fe(CN)_6]$ was dissolved to concentrations of $x$ mM, $x = 0.5, 1, 3, 5$ mM.

Potassium ferricyanide and potassium ferrocyanide form a redox couple in a reversible reaction involving a 1-electron transfer process, from potassium ferricyanide ($K_3[Fe(CN)_6]$) to potassium ferrocyanide ($K_4[Fe(CN)_6]$):

$$K_4[Fe(CN)_6] \leftrightarrow K_3[Fe(CN)_6] + e \qquad (19)$$

The diffusion coefficient of potassium ferricyanide ($D_0$) in our electrolyte was reported as $4.18 \times 10^{-10}$ m$^2$/s [31]. The diffusion coefficient of potassium ferrocyanide ($D_R$) in our electrolyte was obtained following the approach reported in [31] by fitting CV measurements to the Randles–Sevcik equation, yielding $3.56 \times 10^{-10}$ m$^2$/s.

The change in refractive index per unit concentration of potassium ferricyanide ($K_3[Fe(CN)_6]$), $\alpha_O$, is needed to connect the experimental results with theory. The refractive index of our four concentrations of potassium ferricyanide, $C_O = 0.5, 1, 3, 5$ mM (0.35, 0.69, 2.07 and 3.45 mol/m$^3$), dissolved in our electrolyte, were measured using a prism coupler (Metricon) at $\lambda_0 = 1312$ nm and plotted in Fig. 3. Fitting Eq. (3) with $C_R = 0$ to the data yields $\alpha_O = 0.0003$ m$^3$/mol as the slope, and $n_e = 1.3323$ as the intercept ($C_O = 0$). We approximate $\alpha_R$ as $\alpha_O$ because potassium ferricyanide and potassium ferrocyanide are similar molecules, surmised to elicit a similar optical polarizability.

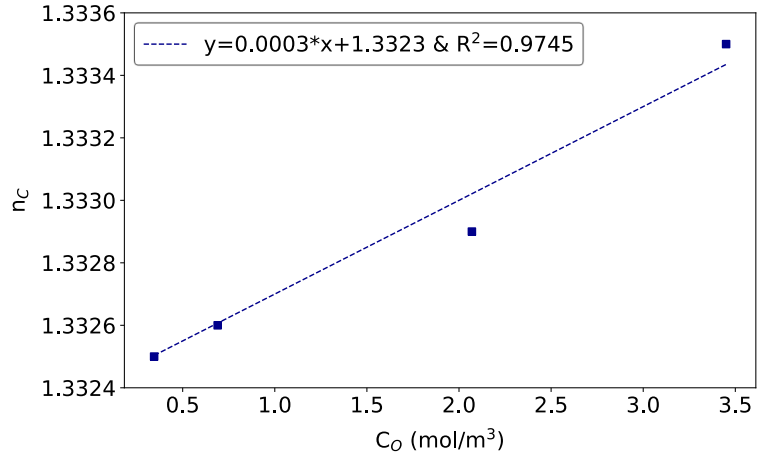

Fig. 3. Refractive index of the electrochemical solution, $n_c$, vs. concentration of potassium ferricyanide $K_3[Fe(CN)_6]$, $C_O$, in our electrolyte (100 mM KNO$_3$ + 158.5 mM glycerol). Best fit of Eq. (3) given in legend.

**Results and Discussion**

*A. Chronoamperometric responses at low optical (SPP) power*

Chronoamperometric measurements were obtained using an on-chip WE and CE, following the experimental scheme sketched in Fig. 1(a) and described in the previous sub-sections. The chronoamperometric responses of four concentrations of potassium ferricyanide ($C_O = 0.5, 1, 3, 5$ mM) in our electrolyte were measured under low incident optical power. The incident optical power was set to $P_{inc} = 0.794$ mW (-1 dBm), which is low enough to ensure that the SPPs propagating along the WE do not affect the electrochemical response and serve only as a probe of electrochemical activity thereon. A forward step potential of from 0 to 450 mV (*vs*. Ag/AgCl) was applied to induce oxidation of the redox species on the WE. The measured current density $j(t)$ is plotted in Fig. 4(a) and the corresponding output optical power, measured simultaneously, $P_{out}(t)$, is plotted in Fig. 4(b). Fig. 4(a) shows a clear dependence of the current responses on the concentration as expected.

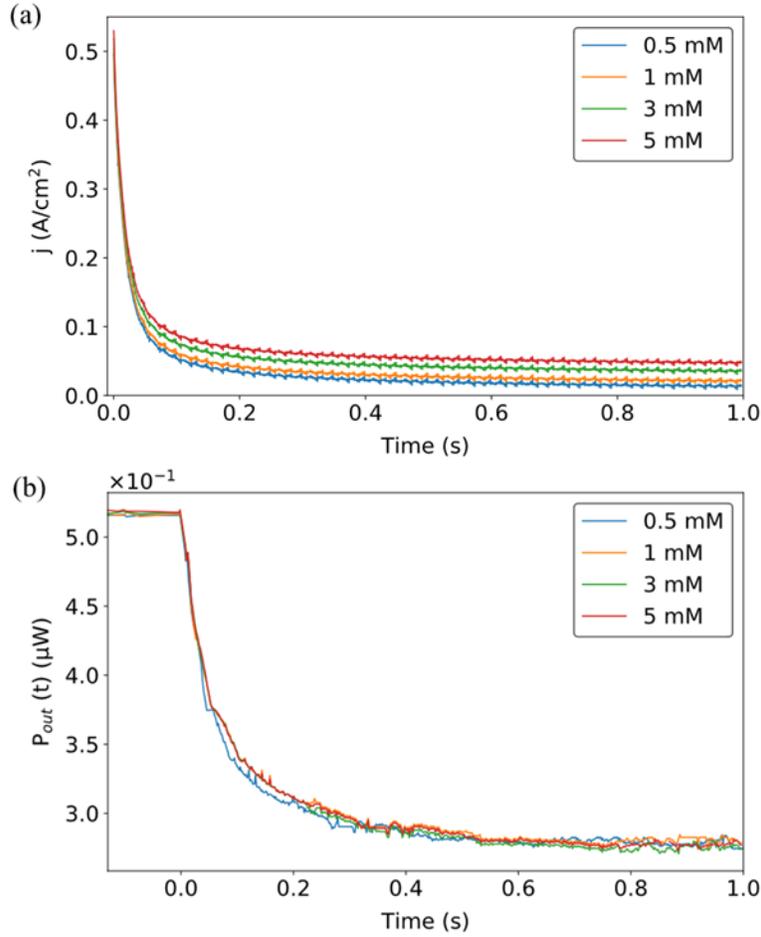

Fig. 4. (a) Chronoamperometric responses on a Au WE to a 450 mV (*vs*. Ag/AgCl) potential step, for different concentrations of potassium ferricyanide $K_3[Fe(CN)_6]$ (legend), in our electrolyte (100 mM KNO3 + 158.5 mM glycerol), under low-power SPP excitation. (b) Measured output optical power at $\lambda_0$ =1350 nm *vs*. time as the chronoamperometric response of Part (a) was recorded.

Using $\alpha_R$, $\alpha_O$, $D_R$ and $D_O$ given in the previous section, with $n = 1$ and $P_{inc} = 0.794$ mW, $S_{B,T_s}$ can be obtained over the time frame of the chronoamperometric responses by using Eq. (18) with the measured responses of $j(t)$ and $P_{out}(t)$ (Figs. 4(a) and 4(b)). In doing so, the time $t = 0$ was taken as the time immediately before application of the step potential (*i.e.*, $P_{out}(0) = P_{out}(0^-)$), and the time convolution of $j(t)$ was computed by numerical integration. Fig. 5 plots $S_{B,T_s}$ over the duration of the experiment for the four concentrations of potassium ferricyanide investigated. $S_{B,T_s}$ is constant over time and over concentration – this is expected because $S_{B,T_s}$ (bulk sensitivity of the sensing channel) is set by design and is independent of the redox species and its concentration, of the electrochemical technique applied, and of any associated electrochemical variables. This result also validates implicitly that output optical

p. 12

power evolves in time as the convolution of the electrochemical current density. The value of $S_{B,T_s}$ averaged over time and over all concentrations is 0.0227 [RIU$^{-1}$], which yields 18.023 [W/RIU] for $S_{B,SPP}$ at $P_{inc} = 0.794$ mW.

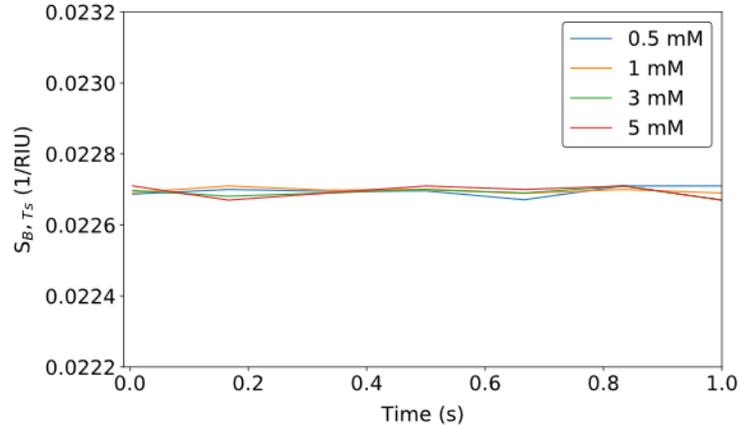

Fig. 5. $S_{B,T_s}$ extracted over the duration of the chronoamperometric responses of Fig. 4 (same legend).

*B. Cyclic voltammetry at low optical (SPP) power*

Cyclic voltammetry (CV) measurements were obtained with 0.5 mM of potassium ferricyanide in electrolyte, at a scan rate of 100 mV/s over a potential window of 0 to 500 mV (*vs*. Ag/AgCl) to drive the reversible reaction of Eq. (19). Fig. 6(a) shows the triangular voltage waveform applied to obtain the CV measurements. Fig. 6(b) shows the measured current density *vs*. time over one period of the drive voltage. The incident optical power was again set to $P_{inc} = 0.794$ mW to probe electrochemical activity on the WE. The corresponding output optical power measured simultaneously is plotted in Fig. 6(c), along with that calculated using Eq. (17) with the current density measured in Fig. 6(b), using numerical integration, $n = 1$, and the coefficients $\alpha_R, \alpha_o, D_R, D_O$ and $S_{B,T_s}$ given in the previous sub-sections. Very good agreement between the measured and computed power responses is noted from Fig. 6(c), explicitly confirming that the power response follows the time convolution of the current density.

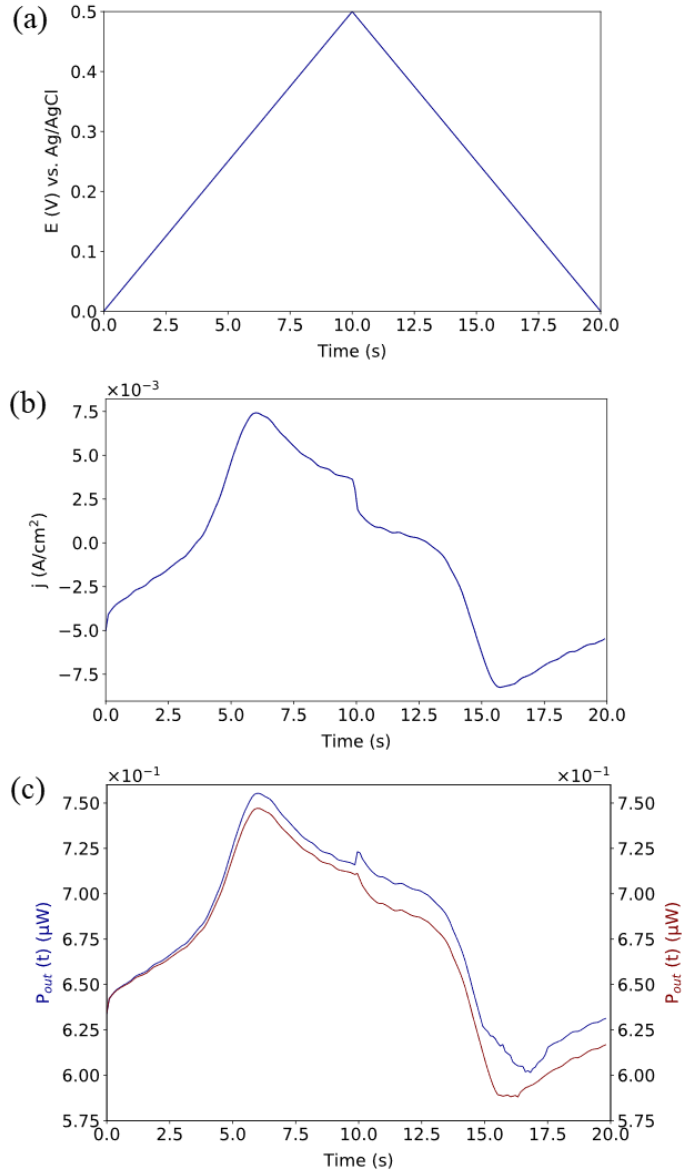

Fig. 6. (a) Triangular voltage waveform applied to obtain CV measurements at a scan speed of 100 mV/s. (b) Current density *vs.* time measured over one voltage period on a Au WE immersed in 0.5 mM $K_3[Fe(CN)_6]$ in our electrolyte (100 mM $KNO_3$ + 158.5 mM glycerol), under low-power SPP excitation. (c) Measured (blue) and computed (red) output optical power responses at $\lambda_0$ =1350 nm, over the same voltage period as in Part (b).

Eq. (17) states that the output optical power does not depend explicitly on electrochemical variables, such as the scan rate in CV measurements. To verify this, CV measurements were obtained for different scan rates (20, 50 and 100 mV/s) while monitoring the output optical power due to SPPs propagating along the WE ($P_{inc}$ = 0.794 mW).

Fig. 7(a) shows the measured cyclic voltammograms, revealing that the peak current density increases by more than 2× when the scan rate increases from 20 to 100 mV/s. The corresponding output optical powers measured simultaneously are plotted as the power voltammograms of Fig. 7(b), along with those computed using Eq. (17) with the measured current densities of Fig. 7(a), revealing good agreement. Essentially no dependence on scan rate is observed in Fig. 7(b), as the measured power voltammograms collapse (approximately) to a single curve. The power voltammograms also confirm that the propagating SPPs probe primarily changes in refractive index between the reduced and oxidised forms of the redox species (potassium ferricyanide).

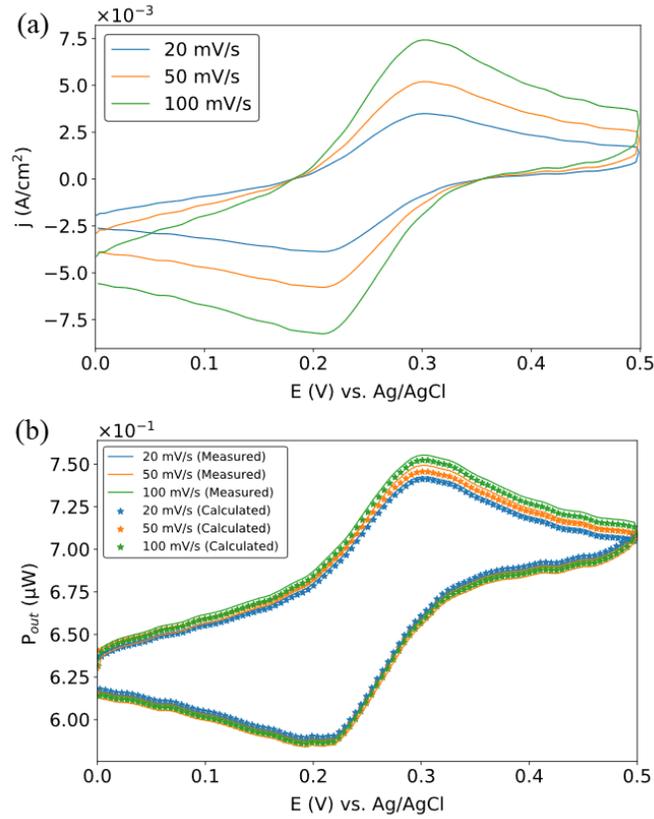

Fig. 7. (a) Current density voltammograms on a Au WE immersed in 0.5 mM of $K_3[Fe(CN)_6]$ in our electrolyte (100 mM $KNO_3$ + 158.5 mM glycerol), at various scan rates (legend), under low-power SPP excitation. (b) Power voltammograms measured simultaneously with the current density voltammograms of Part (a) at $\lambda_0 = 1350$ nm (solid curves). The corresponding computed voltammograms are also plotted (stars).

*C. Cyclic voltammetry at high optical (SPP) power*

CV measurements were then obtained in 0.5 mM of potassium ferricyanide in our electrolyte, at a scan rate of 100 mV/s, with SPPs propagating along the Au WE excited at increasing incident optical powers, $P_{inc}$, such that energetic carriers increasingly affect redox currents, while monitoring the output optical power. In this case, the SPPs not only serve to probe electrochemical activity on the WE, but also to open (pump) new redox channels involving energetic carriers.

The measured cyclic voltammograms are plotted in Fig. 8(a), including a reference curve with the laser off plotted as the black curve. From the data plotted in Fig. 8(a), the peak oxidation and reduction current densities and the potentials corresponding to the peak currents are identified as a function of incident optical power and plotted in Fig. 8(b). From Fig. 8(b) it is noted that the redox current densities increase by up to 10× and the oxidation, reduction and equilibrium potentials drop by up to 2× with increasing SPP excitation (increasing $P_{inc}$) beyond a threshold at $P_{inc} \sim 1$ mW. These changes are due to energetic carriers created along the WE as SPPs propagate and are absorbed therein - energetic electrons transfer more readily than electrons at equilibrium from the WE to the redox species thereby enhancing the reduction current, and energetic holes transfer more readily from the WE to enhance the oxidation current. The threshold at $P_{inc} \sim 1$ mW indicates the opening of these redox channels as the current density associated with energetic carrier transfer overcomes noise in the system and the equilibrium redox current densities. (Extensive and *in-situ* thermal control experiments carried out in previous work on a similar set-up rule out thermal effects as the cause of the current increase [23].)

The corresponding output optical power measured simultaneously is plotted in Fig. 8(c) as power voltammograms. Fig. 8(c) reveals two important features. The first is the power scaling implied by Eq. (17), which states that the convolved signal ($P_{out}$) scales to larger powers with increasing $P_{inc}$, independently of the electrochemical variables – this is observable as voltammograms that rise on the plot with increasing $P_{inc}$. This feature is useful for improving the signal-to-noise ratio of the measurement. The second is that the power voltammograms open as $P_{inc}$ increases – a feature better observed from the normalised voltammograms of Fig. 8(d). The opening is particularly noteworthy beyond threshold ($P_{inc} > 1$ mW), where energetic carrier transfer significantly affects the current density.

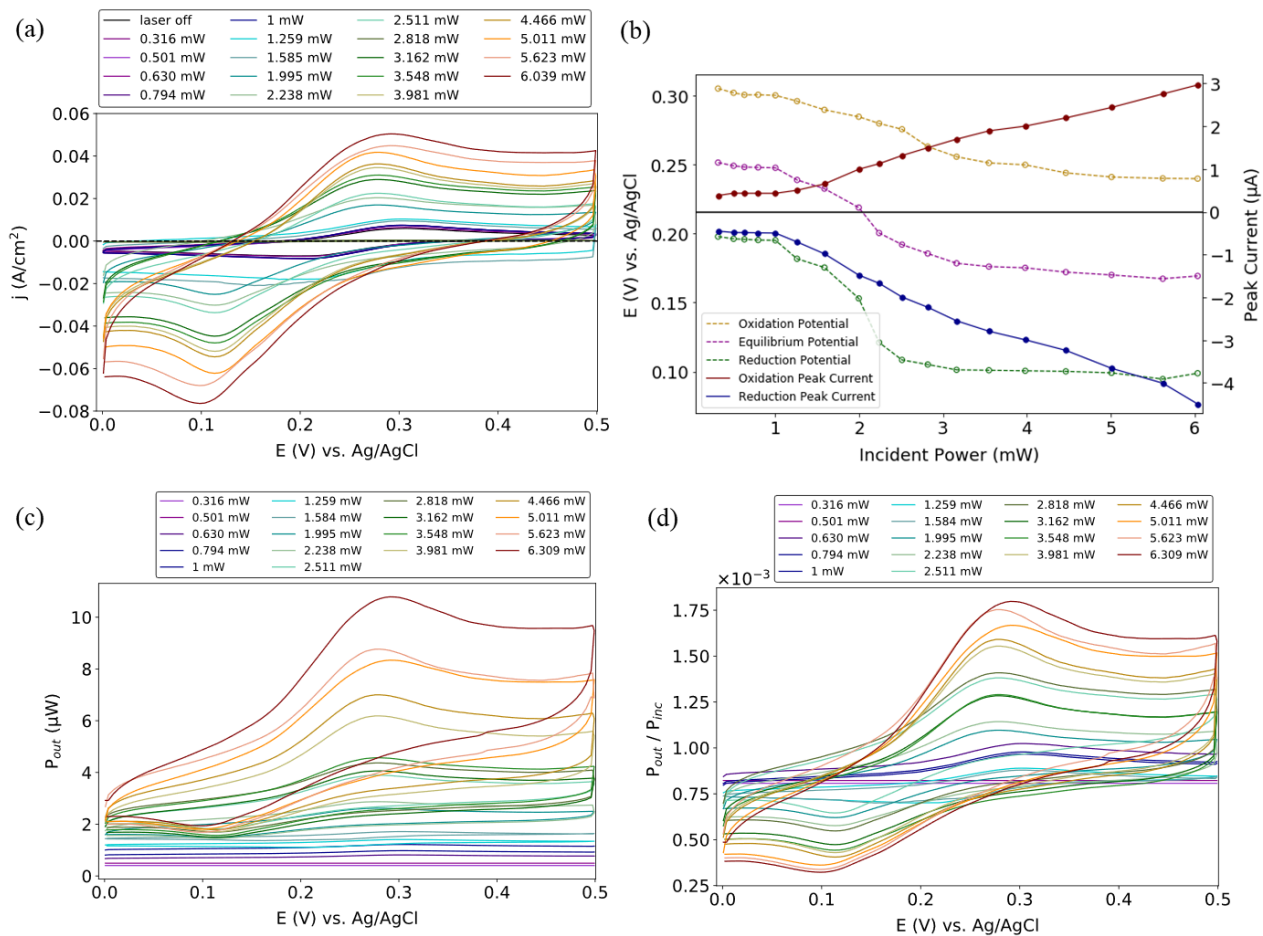

Fig. 8. (a) Current density voltammograms obtained on a Au WE immersed 0.5 mM of $K_3[Fe(CN)_6]$ in our electrolyte (100 mM $KNO_3$ + 158.5 mM glycerol), at a scan rate of 100 mV/s, with the WE supporting propagating SPPs excited at the incident optical powers ($P_{inc}$) indicated in the legend. (b) Peak oxidation and reduction currents and corresponding potentials (oxidation, reduction, equilibrium) *vs*. incident optical power, summarised from Part (a). (c) Power voltammograms measured simultaneously with the current density voltammograms of Part (a), for the incident optical powers ($P_{inc}$) indicated in the legend. (d) Power voltammograms of Part (c), normalised to the incident power, $P_{inc}$.

Fig. 9 compares three of the normalised voltammograms of Fig. 8(d) measured at three values of $P_{inc}$ above threshold ($P_{inc} > 1$ mW), with normalised voltammograms computed with Eq. (17) using the corresponding measured current densities plotted in Fig. 8(a). Excellent agreement is noted, indicating that $P_{out}$ remains proportional to the time convolution of the current density.

p. 17

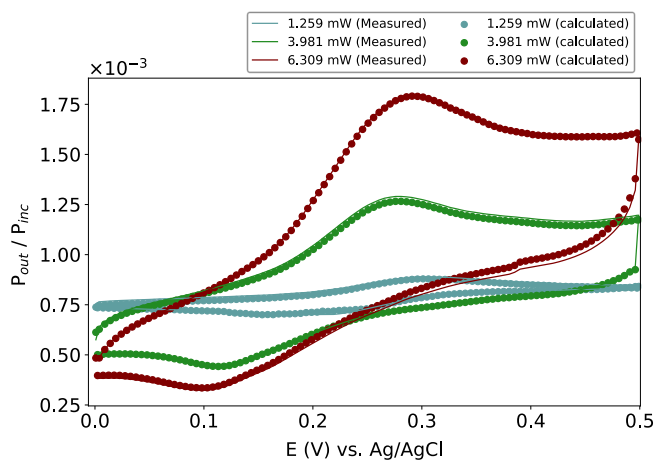

Fig. 9. Normalised power voltammograms beyond threshold, from Fig. 8, compared with corresponding computed voltammograms.

High SPP powers open new redox channels associated with energetic carriers which not only enhance the electrochemical current densities (Fig. 8(a)), but also their convolution as directly measured via the output optical power (Fig. 8(d)). Enhanced convolution indicates that the concentration of electroactive species at the working electrode is also enhanced (Eq. (9)), the result of increased reaction rates which follow enhanced current densities, as energetic carriers transfer more readily to the redox species.

**Conclusions**

We investigated a Au stripe waveguide along which SPPs propagate, operating simultaneously as an electrochemical electrode. Theory relating the real-time response of the output optical power to the time convolution of the current density was developed and validated experimentally via chronoamperometry measurements obtained at different concentrations of the redox species, and cyclic voltammetry measurements obtained at different scan rates. The bulk optical sensitivity of the waveguide WE, $S_{B,T_s}$, was determined by simultaneously fitting the measured optical power and the measured current density from chronoamperometric responses to a theoretical relationship linking them.

At low power, the SPPs propagating along the WE probe changes in refractive index between the reduced and oxidised forms of the redox species. Increasing the SPP power drives redox reactions involving energetic carriers (hot electrons and holes) created by SPP absorption in the Au WE, leading to significantly enhanced redox current densities and convolution signals. Enhanced convolution indicates that the concentrations of electroactive species at the working electrode are also enhanced, resulting from increased reaction rates which follow increased current densities, as energetic carriers (hot electrons and holes) transfer more readily to the redox species.

Thus, exciting SPPs along a working electrode is useful not only as a probe of electrochemical activity, but also to drive significant electrochemical enhancement.